\input harvmac
\input graphicx
\input color

\def\Title#1#2{\rightline{#1}\ifx\answ\bigans\nopagenumbers\pageno0\vskip1in
\else\pageno1\vskip.8in\fi \centerline{\titlefont #2}\vskip .5in}

%
%
\ifx\includegraphics\UnDeFiNeD\message{(NO graphicx.tex, FIGURES WILL BE IGNORED)}
\def\figin#1{\vskip2in}
\else\message{(FIGURES WILL BE INCLUDED)}\def\figin#1{#1}
\fi
\def\Fig#1{Fig.~\the\figno\xdef#1{Fig.~\the\figno}\global\advance\figno
 by1}
%
%
%
%

\font\ticp=cmcsc10

\def \purge#1 {\textcolor{magenta}{#1}}
\def \new#1 {\textcolor{blue}{#1}}
\def\comment#1{}

\def\\{\cr}
\def\text#1{{\rm #1}}
\def\frac#1#2{{#1\over#2}}

\def\calo{{\cal O}}

\def\calh{{\cal H}}

\def\eg{{\it e.g.}}
\def\roughly#1{\mathrel{\raise.3ex\hbox{$#1$\kern-.75em\lower1ex\hbox{$\sim$}}}}
\font\bbbi=msbm10 
\def\mathbb#1{\hbox{\bbbi #1}}

\def\mthsu{\mathsurround=0pt  }
\def\leftrightarrowfill{$\mthsu \mathord\leftarrow\mkern-6mu\cleaders
  \hbox{$\mkern-2mu \mathord- \mkern-2mu$}\hfill
  \mkern-6mu\mathord\rightarrow$}
\def\overleftrightarrow#1{\vbox{\ialign{##\crcr\leftrightarrowfill\crcr\noalign{\kern-1pt\nointerlineskip}$\hfil\displaystyle{#1}\hfil$\crcr}}}
\overfullrule=0pt

%
%
\lref\MaSu{
  J.~Maldacena and L.~Susskind,
  ``Cool horizons for entangled black holes,''
Fortsch.\ Phys.\  {\bf 61}, 781 (2013).
[arXiv:1306.0533 [hep-th]].
}
\lref\NVNLT{
  S.~B.~Giddings,
  ``Modulated Hawking radiation and a nonviolent channel for information release,''
[arXiv:1401.5804 [hep-th]].
}
\lref\BHQIUE{
  S.~B.~Giddings,
  ``Black holes, quantum information, and unitary evolution,''
  Phys.\ Rev.\ D {\bf 85}, 124063 (2012).
[arXiv:1201.1037 [hep-th]].
}
\lref\SGmodels{
  S.~B.~Giddings,
   ``Models for unitary black hole disintegration,''  Phys.\ Rev.\ D {\bf 85}, 044038 (2012)
[arXiv:1108.2015 [hep-th]].
}
\lref\NVNL{
  S.~B.~Giddings,
  ``Nonviolent nonlocality,''
  Phys.\ Rev.\ D {\bf 88},  064023 (2013).
[arXiv:1211.7070 [hep-th]].
}
\lref\Locbdt{
  S.~B.~Giddings and M.~Lippert,
  ``The Information paradox and the locality bound,''
Phys.\ Rev.\ D {\bf 69}, 124019 (2004).
[hep-th/0402073].
}
\lref\BCS{
  N.~Bao, S.~M.~Carroll and A.~Singh,
  ``The Hilbert Space of Quantum Gravity Is Locally Finite-Dimensional,''
Int.\ J.\ Mod.\ Phys.\ D {\bf 26}, no. 12, 1743013 (2017).
[arXiv:1704.00066 [hep-th]].
}
\lref\Chru{Piotr T. Chru\'sciel, ``Anti-gravity \`a la Carlotto-Schoen,"  arXiv:1611.01808 [math.DG].}
\lref\DoGithree{
  W.~Donnelly and S.~B.~Giddings,
  ``How is quantum information localized in gravity?,''
Phys.\ Rev.\ D {\bf 96}, no. 8, 086013 (2017).
[arXiv:1706.03104 [hep-th]].
}
\lref\DoGifour{W.~Donnelly and S.~B.~Giddings, unpublished.
}
\lref\UQM{
  S.~B.~Giddings,
  ``Universal quantum mechanics,''
Phys.\ Rev.\ D {\bf 78}, 084004 (2008).
[arXiv:0711.0757 [quant-ph]].
}
\lref\Haag{R. Haag, {\sl Local quantum physics, fields, particles, algebras}, Springer (Berlin, 1996).}
\lref\SGalg{
  S.~B.~Giddings,
 ``Hilbert space structure in quantum gravity: an algebraic perspective,''
JHEP {\bf 1512}, 099 (2015).
[arXiv:1503.08207 [hep-th]].
}
\lref\locbdi{
  S.~B.~Giddings and M.~Lippert,
  ``Precursors, black holes, and a locality bound,''
Phys.\ Rev.\ D {\bf 65}, 024006 (2002).
[hep-th/0103231].
}
\lref\DoGione{
  W.~Donnelly and S.~B.~Giddings,
  ``Diffeomorphism-invariant observables and their nonlocal algebra,''
Phys.\ Rev.\ D {\bf 93}, no. 2, 024030 (2016), Erratum: [Phys.\ Rev.\ D {\bf 94}, no. 2, 029903 (2016)].
[arXiv:1507.07921 [hep-th]].
}
\lref\DoGitwo{
  W.~Donnelly and S.~B.~Giddings,
  ``Observables, gravitational dressing, and obstructions to locality and subsystems,''
Phys.\ Rev.\ D {\bf 94}, no. 10, 104038 (2016).
[arXiv:1607.01025 [hep-th]].
}
\lref\HPS{
  S.~W.~Hawking, M.~J.~Perry and A.~Strominger,
  ``Soft Hair on Black Holes,''
Phys.\ Rev.\ Lett.\  {\bf 116}, no. 23, 231301 (2016).
[arXiv:1601.00921 [hep-th]]\semi
``Superrotation Charge and Supertranslation Hair on Black Holes,''
JHEP {\bf 1705}, 161 (2017).
[arXiv:1611.09175 [hep-th]].
}
\lref\CCM{
  C.~Cao, S.~M.~Carroll and S.~Michalakis,
  ``Space from Hilbert Space: Recovering Geometry from Bulk Entanglement,''
Phys.\ Rev.\ D {\bf 95}, no. 2, 024031 (2017).
[arXiv:1606.08444 [hep-th]].
}
\lref\CaCa{
  C.~Cao and S.~M.~Carroll,
  ``Bulk Entanglement Gravity without a Boundary: Towards Finding Einstein's Equation in Hilbert Space,''
[arXiv:1712.02803 [hep-th]].
}
\lref\CaSi{
  S.~M.~Carroll and A.~Singh,
  ``Mad-Dog Everettianism: Quantum Mechanics at Its Most Minimal,''
[arXiv:1801.08132 [quant-ph]].
}
\lref\Hartone{
  J.~B.~Hartle,
  ``The Quantum mechanics of cosmology,''
  in {\sl Quantum cosmology and baby universes : proceedings}, 7th Jerusalem Winter School for Theoretical Physics, Jerusalem, Israel, December 1989, ed. S. Coleman, J. Hartle, T. Piran, and S. Weinberg (World Scientific, 1991).
}
\lref\Harttwo{
  J.~B.~Hartle,
  ``Space-time coarse grainings in nonrelativistic quantum mechanics,''
  Phys.\ Rev.\  D {\bf 44}, 3173 (1991).
}
\lref\HartLH{
  J.~B.~Hartle,
  ``Space-Time Quantum Mechanics And The Quantum Mechanics Of Space-Time,''
  arXiv:gr-qc/9304006.
}
\lref\HartPuri{
  J.~B.~Hartle,
  ``Quantum Mechanics At The Planck Scale,''
  arXiv:gr-qc/9508023.
}
\lref\Eins{A. Einstein, ``Quanten-Mechanik und Wirklichkeit," Dialectica {\bf 2} (1948) 320.}
\lref\Howa{D. Howard, ``Einstein on locality and separability," Studies in History and Philosophy of Science Part A  {\bf 16} no. 3, (1985)
171.}
\lref\vanR{
  M.~Van Raamsdonk,
 ``Building up spacetime with quantum entanglement,''
Gen.\ Rel.\ Grav.\  {\bf 42}, 2323 (2010), [Int.\ J.\ Mod.\ Phys.\ D {\bf 19}, 2429 (2010)].
[arXiv:1005.3035 [hep-th]].
}
\lref\GiRo{
  S.~B.~Giddings and M.~Rota,
 ``Quantum information/entanglement transfer rates between subsystems,''
[arXiv:1710.00005 [quant-ph]].
}
\lref\BRSSZ{
  A.~R.~Brown, D.~A.~Roberts, L.~Susskind, B.~Swingle and Y.~Zhao,
  ``Complexity, action, and black holes,''
Phys.\ Rev.\ D {\bf 93}, no. 8, 086006 (2016).
[arXiv:1512.04993 [hep-th]].
}
\lref\Sussfall{
  L.~Susskind,
  ``Why do Things Fall?,''
[arXiv:1802.01198 [hep-th]].
}
\lref\NPGNL{
  S.~B.~Giddings,
  ``(Non)perturbative gravity, nonlocality, and nice slices,''
Phys.\ Rev.\ D {\bf 74}, 106009 (2006).
[hep-th/0606146].
}
\lref\Mald{
  J.~M.~Maldacena,
  ``Eternal black holes in anti-de Sitter,''
JHEP {\bf 0304}, 021 (2003).
[hep-th/0106112].
}
\lref\Heem{
  I.~Heemskerk,
  ``Construction of Bulk Fields with Gauge Redundancy,''
JHEP {\bf 1209}, 106 (2012).
[arXiv:1201.3666 [hep-th]].
}
\lref\KaLigrav{
  D.~Kabat and G.~Lifschytz,
  ``Decoding the hologram: Scalar fields interacting with gravity,''
Phys.\ Rev.\ D {\bf 89}, no. 6, 066010 (2014).
[arXiv:1311.3020 [hep-th]].
}
\lref\Zure{W. H. Zurek, ``Quantum darwinism, classical reality, and the 
randomness of quantum jumps," Physics Today {\bf 67} 44, 	arXiv:1412.5206.
}
\lref\LQGST{
  S.~B.~Giddings,
 ``Locality in quantum gravity and string theory,''
Phys.\ Rev.\ D {\bf 74}, 106006 (2006).
[hep-th/0604072].
}
\lref\GiKi{
  S.~B.~Giddings and A.~Kinsella,
  ``Gauge-invariant observables, gravitational dressings, and holography in AdS,''
[arXiv:1802.01602 [hep-th]].
}
\lref\BuVe{
  D.~Buchholz and R.~Verch,
  ``Scaling algebras and renormalization group in algebraic quantum field theory,''
Rev.\ Math.\ Phys.\  {\bf 7}, 1195 (1995).
[hep-th/9501063].
}
\lref\Yngv{
  J.~Yngvason,
  ``The Role of type III factors in quantum field theory,''
Rept.\ Math.\ Phys.\  {\bf 55}, 135 (2005).
[math-ph/0411058].
}
\lref\ZLL{
  P.~Zanardi, D.~A.~Lidar and S.~Lloyd,
 ``Quantum tensor product structures are observable induced,''
Phys.\ Rev.\ Lett.\  {\bf 92}, 060402 (2004).
[quant-ph/0308043].
}
\lref\Harl{
  D.~Harlow,
  ``Wormholes, Emergent Gauge Fields, and the Weak Gravity Conjecture,''
JHEP {\bf 1601}, 122 (2016).
[arXiv:1510.07911 [hep-th]].
}
\lref\DoFr{
  W.~Donnelly and L.~Freidel,
  ``Local subsystems in gauge theory and gravity,''
JHEP {\bf 1609}, 102 (2016).
[arXiv:1601.04744 [hep-th]].
}
\lref\GuJa{
  M.~Guica and D.~L.~Jafferis,
  ``On the construction of charged operators inside an eternal black hole,''
SciPost Phys.\  {\bf 3}, no. 2, 016 (2017).
[arXiv:1511.05627 [hep-th]].
}
\lref\CPR{
  J.~S.~Cotler, G.~R.~Penington and D.~H.~Ranard,
  ``Locality from the Spectrum,''
[arXiv:1702.06142 [quant-ph]].
}
\lref\PaRa{
  K.~Papadodimas and S.~Raju,
  ``Local Operators in the Eternal Black Hole,''
Phys.\ Rev.\ Lett.\  {\bf 115}, no. 21, 211601 (2015).
[arXiv:1502.06692 [hep-th]].
}
\lref\CoSc{
  J.~Corvino and R.~M.~Schoen,
  ``On the asymptotics for the vacuum Einstein constraint equations,''
J.\ Diff.\ Geom.\  {\bf 73}, no. 2, 185 (2006).
[gr-qc/0301071].
}
\lref\ChDe{
  P.~T.~Chrusciel and E.~Delay,
  ``On mapping properties of the general relativistic constraints operator in weighted function spaces, with applications,''
Mem.\ Soc.\ Math.\ France {\bf 94}, 1 (2003).
[gr-qc/0301073].
}
\lref\NVNLpost{
  S.~B.~Giddings,
  ``Nonviolent unitarization: basic postulates to soft quantum structure of black holes,''
JHEP {\bf 1712}, 047 (2017).
[arXiv:1701.08765 [hep-th]].
}
\lref\HSTone{
  T.~Banks and W.~Fischler,
  ``M theory observables for cosmological space-times,''
[hep-th/0102077].
}
\lref\HSTrev{
  T.~Banks,
  ``Lectures on Holographic Space Time,''
[arXiv:1311.0755 [hep-th]].
}
\lref\BGH{
  A.~Bzowski, A.~Gnecchi and T.~Hertog,
  ``Interactions resolve state-dependence in a toy-model of AdS black holes,''
[arXiv:1802.02580 [hep-th]].
}
\Title{
\vbox{\baselineskip12pt  
}}
{\vbox{\centerline{Quantum-first gravity
} }}

\centerline{{\ticp 
Steven B. Giddings\footnote{$^\ast$}{Email address: giddings@ucsb.edu}
} }
\centerline{\sl Department of Physics}
\centerline{\sl University of California}
\centerline{\sl Santa Barbara, CA 93106}
\vskip.10in
\centerline{\bf Abstract}

This paper elaborates on an intrinsically quantum 
approach to gravity, which begins with a general framework for quantum mechanics and then 
seeks to identify additional mathematical structure on Hilbert space that is responsible for gravity and other phenomena.  A key principle in this approach is that of correspondence: this structure should reproduce spacetime, general relativity, and quantum field theory in a limit of weak gravitational fields.  A central question is that of ``Einstein separability," and asks how to define mutually independent subsystems, {\it e.g.} through localization.  Familiar definitions involving tensor products or operator subalgebras do not clearly accomplish this in gravity, as is seen in the correspondence limit.  Instead, gravitational behavior, particularly gauge invariance, suggests a network of Hilbert subspaces related via inclusion maps, contrasting with other approaches based on tensor-factorized Hilbert spaces.  Any such localization structure is also expected to place strong constraints on evolution, which are also supplemented by the constraint of unitarity.

\Date{}

\newsec{The quantum mechanics first approach}

A profound challenge for modern fundamental physics is to find a theory of gravity consistent with quantum reality.  Various attempts to {\it quantize} gravity have met with serious difficulties. But, an alternate approach is to begin with the goal in mind and consider what structure an intrinsically quantum-mechanical theory should have in order to describe gravity.  Indeed, the rigidity of quantum mechanics suggests that the need for a theory to fit within quantum mechanics could be tightly constraining, and thus provide a new direction which is supplemented by other important clues that we already have about the quantum nature of gravity.

Such a ``quantum-first" approach to gravity has been advocated in \refs{\UQM\BHQIUE-\SGalg}, and also recently by Carroll and collaborators\refs{\CCM\CaCa-\CaSi}, and will be elaborated further here.\foot{For an earlier but somewhat different approach to describing quantum structure for gravity, see \refs{\HSTone,\HSTrev}.}  In contrast, previous approaches have started with a classical theory, such as general relativity (GR), and attempted to apply a set of rules to quantize the theory.  These approaches have encountered vexing problems: at first that of non-renormalizability, but perhaps more profoundly, also the problem of respecting the basic quantum principle of unitarity, when one accounts for black hole formation and decay.  Similar comments can be made regarding attempts to quantize strings; this seems to avoid non-renormalizability but still has not resolved the ``unitarity crisis" associated with black holes.
The troubles with attempts to {\it quantize} gravity suggest taking a different tack -- that of ``geometrizing" quantum mechanics.

While it is certainly conceivable that we would have to alter quantum mechanics to describe gravity, a worthy and well-motivated goal is  thus to investigate whether gravity can be described in an inherently quantum-mechanical framework.  (If that is not possible, this is also critical information.)  A starting point here is a suitably general framework for quantum mechanics. For example, ref.~\refs{\Hartone\Harttwo\HartLH-\HartPuri} proposed a ``generalized quantum mechanics,"  but this is still too closely tied to the notion of quantizing a classical theory and is not sufficiently general.  This motivated a proposal\UQM\ for the essential postulates of quantum mechanics -- those of ``universal quantum mechanics"  (UQM).  In brief, as outlined in \UQM, these are the existence of a linear space of states with an inner product (``Hilbert space"), and the existence of linear hermitian operators that are interpreted as providing quantum observables; the postulates include unitarity, {\it e.g.} the S-matrix, in appropriate circumstances.  Of course, additional structure needs to be furnished to describe a complete and specific quantum theory.

Such a quantum-first approach has precedent in one approach to understanding local quantum field theory (LQFT).  LQFT can be regarded as the solution to a problem, that of reconciling the postulates of quantum mechanics with the additional  postulates of special relativity and of  spacetime locality.  Specifically, one seeks to describe a Hilbert space with a privileged role for local observables, and with an action of the Poincar\'e group.  Defining local operators requires introducing additional structure, that of Minkowski space.  Part of the lore taught in some LQFT classes and books is that LQFT, \eg\ via the introduction of Fock space, is the unique way to build a quantum-mechanical theory implementing these additional principles.  The key principle of locality is encoded in the fact that local operators commute outside each-other's light cones, an important structural aspect to which we will return.

In seeking an intrinsically quantum description of gravity, an important guide is therefore that of needing to provide an additional {\it mathematically consistent structure} within  general principles of quantum mechanics.  This clearly is not enough guidance, but is supplemented by another key principle\refs{\NPGNL,\SGalg}: that of {\it correspondence}.  Specifically, in weak gravity regimes we know that a more fundamental theory must match onto LQFT together with weak gravitational phenomena which may be treated in an expansion in the gravitational coupling $G$ that matches perturbative GR; there is abundant evidence for this in current experimental physics.  

The need for a mathematical structure within UQM {\it and} to satisfy correspondence provide a tight set of constraints -- hopefully not too tight to be reconciled.  These provide twin guiding lights in approaching quantum-first gravity.  

A key question, then, is what kind of mathematical structure is needed on Hilbert space, to satisfy these requirements in an economical fashion.  And, a first critical ingredient in any such structure appears to be a basic definition of subsystems of the quantum system.  The importance to physics of a division into subsystems was recognized by no less than Einstein, who wrote\foot{Ref.~\refs{\Eins}; for translation of relevant passages see \refs{\Howa}.}  ``Further, it appears to be essential for this arrangement of the things introduced in physics that, at a specific time, these things claim an existence independent of one another, insofar as these things `lie in different parts of space.'  Without such an assumption of the mutually independent existence (the `being-thus') of spatially distant things, an assumption which originates in everyday thought, physical thought in the sense familiar to us would not be possible.  Nor does one see how physical laws could be formulated and tested without such a clean separation."  

This notion of separability or independence, implemented through a subsystem division, appears crucial, and despite being rather subtle in gravity, is simply assumed in various current approaches to physics.  For example, one proposal is that spacetime ``emerges from entanglement\refs{\vanR,\MaSu}."  However, a notion of entanglement relies, first, on a notion of division of a system into subsystems; it may be easily illustrated that for a given quantum state, different such divisions lead to either zero or nonzero entanglement.  Related comments apply to entropies based on entanglement. Transfer of information, or entanglement (see, \eg, \refs{\GiRo}) also requires division into subsystems between which transfer occurs.  Or, it has been argued that complexity plays a key role in gravity\refs{\BRSSZ}.  However, as we will discuss, definition of such complexity is relative to a subsystem division.  More generally, other attempts to give a set of postulates for quantum mechanics often begin with a factorization postulate associated with the definition of subsystems -- see {\it e.g.} postulate zero of \refs{\Zure}.

In Einstein's description, separability arises from spatial separation, and indeed this gives the basis for a definition of subsystems in LQFT.  Specifically, in LQFT, a natural notion of subsystem arises from networks of subalgebras of observables associated with spacelike-separated neighborhoods.\foot{For further discussion see \refs{\Haag,\SGalg}.  Concretely, examples of operators in the subalgebra are field operators smeared against test functions with support restricted to the neighborhood.}  As described above, such subalgebras of operators will commute, and these subalgebras may be naturally thought of as defining subsystems, \eg\ because these operators may  act to ``create particles" in an independent fashion.  Thus, the underlying spacetime manifold and the locality that it induces plays a key role in defining separability and subsystems in LQFT.  

The key question for quantum-first gravity is then what analogous mathematical structure on Hilbert space allows description of separability and subsystems.  As we will review below, locality commutativity of observables fails\refs{\DoGione,\DoGitwo} in a theory respecting the correspondence principle with weak-field (or classical) GR;\foot{As described below, this also connects with proposals for the importance of ``soft quantum hair" on black holes \refs{\HPS}.}   ironically, in a quantum treatment of Einstein's gravity, Einstein's reliance on locality fails.  
The even more familiar notion of Hilbert space factorization is also problematic, and so other structure is apparently needed.

If such a structure is found, providing a network of subsystems, then that plausibly serves as a quantum {\it replacement} for the role of classical spacetime, which then no longer exists as a precise concept, in the theory.  Instead of starting with a classical geometry and quantizing, we instead begin with quantum mechanics (Hilbert space) and attempt to find correct structure that approximately reduces to spacetime in the correspondence limit.
The foundation provided by such a ``gravitational substrate" on the Hilbert space is expected to be a key element in the formulation of the theory, and of course to have important implications -- as does spacetime locality in LQFT -- for other aspects of the theory such as quantum evolution.

\newsec{Gravitational subsystems and constraints from correspondence}

We face the problem of giving a mathematical description of independent subsystems that is valid in a theory which matches onto LQFT plus perturbative weak-field GR in the correspondence limit.  This correspondence plays an important role.  Specifically, whatever structure is present in the fundamental theory should be approximately present in the perturbative limit, and conversely, known aspects of the perturbative limit should correspond to approximations of the structure, {\it e.g.} in a perturbative expansion, of the more fundamental theory.  And already, in this perturbative expansion, gravity indicates novel mathematical structure.

To see this, we first consider in more detail how subsystems are defined in other familiar theories.  Of course in finite quantum systems, or locally finite ones such as a lattice theory, subsystems are defined in terms of a tensor factorization of the Hilbert space, {\it e.g.} corresponding to degrees of freedom at different lattice sites.  This structure is, however, at odds with Lorentz invariance, and this problem is expressed in the statement that the von Neumann algebras of observables that one encounters in field theory are type III.\foot{See \BuVe, and for review \Yngv.}  Physically this means that there is a problem of infinite entanglement of degrees of freedom if we try to define subsystems separated by a spatial boundary; Lorentz invariance implies that there are entangled degrees of freedom at all scales, and this manifests itself for example in divergences in the corresponding von Neumann entropy of a region.  

This obstacle to factorization explains the preceding statement that subsystems in LQFT should instead be thought of as defined via commuting subalgebras\refs{\Haag,\SGalg}.  In fact, these subalgebras form a ``net," which closely mirrors the structure of the underlying spacetime manifold\Haag.  Specifically, to each open spacetime neighborhood is associated a subalgebra, with those associated to spacelike-separated neighborhoods commuting.  Moreover, these subalgebras properly combine when we take a union of neighborhoods, or restrict if we consider an intersection of neighborhoods, reflecting the topological structure.  This provides a foundation for localization of information in LQFT.

Even weak gravity provides new obstructions to such a definition.  A primary one arises from gauge invariance, which is diffeomorphism invariance in the correspondence limit, and the statement that observables must be gauge invariant.  Local operators are not gauge invariant, since diffeomorphisms move them to a different location.  A local operator at $G=0$ can however be perturbatively promoted to a gauge-invariant operator\DoGione\foot{For previous related work, see \refs{\Heem,\KaLigrav}.} in an expansion in $\kappa=\sqrt{32\pi G}$, but the ``dressing theorem" of \DoGitwo\ shows that this operator must have non-trivial gravitational dressing extending to infinity.  These operators can be determined by the condition that they commute with the gravitational constraints.  Colloquially, field operators create particles, but in gravity a particle is inseparable from its gravitational field extending to infinity.

As a consequence, one perturbatively has dressed operators $\Phi(x)$ associated to a location, but in general\refs{\locbdi,\Locbdt,\DoGione}
\eqn\phicom{[\Phi(x),\Phi(y)]\neq0}
for spacelike $x-y$.  The necessary dressing obstructs an algebraic definition of localized subsystems.  In principle, the gravitational field at the distance of the Andromeda galaxy must change if even one atom is added to Earth.  The locality of LQFT is an idealization that is no longer true in gravity, and an important question is what replaces it.

Since at leading order in $\kappa$ the dressing may be localized to a thin ``gravitational line" extending to infinity\DoGione, one might think that it is possible to at least algebraically describe subsystems associated to narrow neighborhoods extending to infinity.  But, even this appears problematic, when one works to higher order in $\kappa$.\foot{Note, however, that one regularization of a gravitational line is to smear it over a cone extending to infinity, and moreover that in classical GR one can show that even at higher-orders gravitational fields can be localized to conical regions as one approaches infinity\refs{\Chru}.  This provides evidence that such gravitational dressings can be consistent at higher orders in $\kappa$.} The reason for this\SGalg\ is that $[\Phi(x)]^N$ (or a regularized version) creates $N$ particles, with $N$ times the energy of one particle.  At the nonlinear level, we expect that as $N$ grows, the operator must create a gravitational field in an ever-larger region; colloquially, in the limit $N\rightarrow\infty$, it creates an infinitely large black hole, escaping any region to which it was initially restricted.

In fact, we have a related expectation of the non-perturbative gravity theory, namely that there are a finite number of allowed quantum states in a given region, and that this number scales with an area surrounding the region. That is another way of saying why $[\Phi(x)]^N|0\rangle$, where $|0\rangle$ is the vacuum, must be associated to a growing region as $N$ increases.  Of course the fact that the non-perturbative theory is expected to have only finitely many states associated to a region na\"\i vely suggests that one has returned to the locally-finite case, where subsystems correspond to tensor factorizations (see \eg\ \refs{\BCS}), but closer examination of gravitational properties suggest a more subtle gravitational regularization of locally infinite behavior such that this is not precisely the case.

In short, it is not clear that gravitational subsystems may be defined by either tensor products or commuting subalgebras -- gravity indicates different structure, which begins to appear even at leading order in $\kappa$.

So, we can ask whether there is any sensible definition of a ``localized subsystem" in gravity.  More colloquially, we would like to understand if there are localized gravitational qubits, that correspond to non-trivial information (\eg\ ``up" or ``down") that is inaccessible outside a region, or more generally, is inaccessible to a given class of operators.  Initially, we address this at the perturbative level.

A key point in addressing this is to notice that while diffeomorphism-invariant operators must be gravitationally dressed, there is a wide latitude in {\it how} they are gravitationally dressed.  The main constraint, in the example of an asymptotically flat situation, comes from the need for the gravitational dressing to express the Poincar\'e charges of the matter configuration.  This suggests that if different matter configurations can be found in a given region  with the same Poinca\'re charges, then there are consistent gravitational dressings for these such that they cannot be distinguished outside the region.

First, let's understand such structure at $\kappa=0$, in LQFT.  This can be described in terms of a {\it splitting}.\foot{See, {\it e.g.}, \Haag, and references therin.}  Specifically, given a neighborhood $U$ and a bigger ``$\epsilon$-extended" neighborhood $U_\epsilon$, it is a general result that in LQFT one can find a split state $|U_\epsilon\rangle$ so that for operators $A$ and $B$ contained in the subalgebras associated to $U$ and the complement region $U_\epsilon'$, respectively, 
\eqn\spliteq{ \langle U_\epsilon| A B |U_\epsilon\rangle = \langle 0| A|0\rangle      \langle 0|B|0\rangle\ ;}
that is, the state  $|U_\epsilon\rangle$ removes correlations between operators inside $U$ and outside $U_\epsilon$.

We would like a similar structure for $\kappa\neq0$, but the dressing prevents it from being identical.  Specifically, if we dress $A$ to give $\hat A$, then the dressing extends outside $U_\epsilon$ and can be detected by operators in $U_\epsilon'$.  But, in seeking a notion of independent subsystem, we can ask whether there can be different operators $\hat A_\alpha$ with the {\it same} dressing outside $U_\epsilon$, so that the states $\hat A_\alpha |\hat U_\epsilon\rangle$ (with the split state also dressed) are indistinguishable outside $U_\epsilon$.  If so, then these states are good candidates for the states of a localized subsystem, and address the need for independent subsystems via a localization similar to that described by Einstein.

The limitations on localization of algebras suggest that we instead focus on states.  We thus seek a Hilbert subspace of states $\calh^i_{U_\epsilon}$ so that for two states $|\psi,U_\epsilon \rangle,\ |\tilde \psi,U_\epsilon \rangle\in \calh^i_{U_\epsilon}$ and any operator $B$ localized outside $U_\epsilon$, 
\eqn\splstr{ \langle \tilde \psi,U_\epsilon|  B|\psi,U_\epsilon \rangle = \langle \tilde \psi,U_\epsilon| \psi,U_\epsilon \rangle \langle i|B| i\rangle\ ,}
where the matrix element of $B$ only depends on the Hilbert space label $i$.  This means $B$ cannot distinguish the different states in $\calh^i_{U_\epsilon}$, which thus give a generalized notion of localized gravitational qubits.  Such a definition of a ``gravitational split structure" was given in \refs{\DoGithree}, and also has a gauge-theory analog.

The question of whether it is possible to find states and dressings with such split structure was preliminarily addressed in \DoGithree, beginning at the classical level.  Specifically, for a given classical matter distribution in $U$, it was shown that one may choose the gravitational field to be of a standard form outside $U$, that just depends on the total Poincar\'e charges $P_\mu$, $M_{\mu \nu}$ of the distribution.  This standard field could be taken to be a gravitational line, as in \DoGione, or alternately as a linearization of the Kerr solution, together with a boost.\foot{Here we make contact with a generalization of the Corvino-Schoen gluing theorem\refs{\CoSc,\ChDe} to the case with sources\DoGithree.  This theorem states that given initial vacuum initial data, one may find new initial data that agrees with the original data in a compact region, but matches a boosted Kerr solution outside large enough radius.}

Unpublished work\refs{\DoGifour} has begun to extend this to the quantum level.  In particular, one may show that, to linear order in $\kappa$, given a state 
$|\psi,U_\epsilon\rangle$ localized to $U_\epsilon$, it may be provided a ``standard dressing" so that states with identical matrix elements $\langle \tilde \psi,U_\epsilon|  P_\mu |\psi,U_\epsilon \rangle$, $\langle \tilde \psi,U_\epsilon|  M_{\mu\nu} |\psi,U_\epsilon \rangle$ are indistinguishable to linear operators in the metric perturbation from flat space, localized outside $U_\epsilon$.  Specifically, with the perturbation defined as $g_{\mu\nu}= \eta_{\mu\nu}+\kappa h_{\mu\nu}$, and with $y\in U_\epsilon'$
\eqn\grone{\langle \tilde \psi,U_\epsilon|  h_{\mu\nu}(y) |\psi,U_\epsilon\rangle =  \langle \tilde \psi,U_\epsilon|  P_\lambda |\psi,U_\epsilon\rangle\, {\tilde h}_{\mu\nu}^\lambda(y)+ 
 \langle \tilde \psi,U_\epsilon|  M_{\lambda\sigma} |\psi,U_\epsilon\rangle\, {\tilde h}_{\mu\nu}^{\lambda\sigma}(y)}
for some standard gravitational fields ${\tilde h}_{\mu\nu}^\lambda(y)$ and ${\tilde h}_{\mu\nu}^{\lambda\sigma}(y)$ which may be taken to be line-like or Kerr.

A conjecture is that this structure extends to more general non-linear operators localized outside $U_\epsilon$, at least perturbatively, yielding a split gravitational structure \splstr\ associated to $U_\epsilon$.\foot{For gauge invariance, these operators should also be dressed; {\it e.g.} analogues to $\Phi(x)$ may be used.}  Notice that localized matter states cannot be taken to be eigenstates of $P_\mu$ and $M_{\mu\nu}$, since these generate translations and Lorentz transformations which will not leave such a distribution invariant; moreover, in the quantum theory these generators do not commute.  This means that in general these localized states have nontrivial distributions of momentum and angular momentum/boost charge; the goal would be to divide these into equivalence classes with the same matrix elements for non-linear operators in $h_{\mu\nu}$ that are localized outside $U_\epsilon$.  It remains to be seen if this is possible; if not that indicates a more subtle and non-trivial form of gravitational nonlocalization, at higher order in $\kappa$.

So far we have just given an approach to defining a division of one subsystem from the bigger system.  However, our ultimate goal is a mathematical structure on Hilbert space that describes the network of all possible subsystems.  We have reviewed the statement that in LQFT the net of subsystems arising from operator subagebras mirrors the structure of the spacetime manifold. So, in quantum-first gravity, identification of an analogous network provides a candidate for the gravitational substrate, which is the quantum structure replacing spacetime for quantum gravity.

Some of the structure of such a network, which has features implied by behavior of gravity, can be briefly described, though more complete exploration is left for later work.  First, a Hilbert space $\calh^i_{U_\epsilon}$ of states indistinguishable outside $U_\epsilon$ is naturally mapped by inclusion into the full Hilbert space,
\eqn\inclus{\calh^i_{U_\epsilon}\hookrightarrow \calh\ .}
Moreover, if $U_\epsilon$ is contained in a larger neighborhood, $U_\epsilon \subset U_\epsilon'$, we expect an inclusion
\eqn\subu{\calh^{i}_{U_\epsilon}\hookrightarrow\calh^{i\prime}_{U_\epsilon'}\ .}
However, there should not be an $\calh^i_{U_\epsilon}$ for each $U_\epsilon$.  For example, if we considered a neighborhood that was smaller than the Planck length $\sim \kappa$ ({\it e.g.} in the original background metric), then any non-trivial states in this neighborhood have energies $E>1/\kappa$ and thus produce a strong gravitational field extending well beyond the neighborhood.  This is one of the ways that gravity leads to a coarser structure than that of the spacetime.

Likewise, for a larger $U_\epsilon$, not all na\"\i vely-allowed labels $i$ are in fact allowed.  For example, if we consider states of LQFT in $U_\epsilon$ with energy $\langle E\rangle$, there is a corresponding Schwarzschild radius $R(\langle E\rangle)=2G\langle E\rangle$ (with obvious generalization to the higher-dimensional case), and energies such that this radius exceeds the size of the neighborhood lead to strong gravitational fields extending beyond the neighborhood.\foot{For the ultra-boosted case, this condition must be appropriately modified.}    A parameterization of this kind of limitation is the locality bound of \refs{\locbdi,\Locbdt,\LQGST}, which states when the LQFT description of states together with perturbative gravity fails, namely when a state is confined to a region with size $r\roughly<R(\langle E\rangle)$.\foot{Note that, unlike entropic bounds, this bound involves energy localized in a region.}  Indeed, \DoGione\ explicitly checked that in the two-particle case, this bound parameterizes when the violation of commutativity, \phicom, becomes $\calo(1)$, signaling the failure of the LQFT description. 

Similarly if we have a separated pair of neighborhoods $U_\epsilon$, $U_\epsilon'$, we expect an inclusion map of the form
\eqn\prodh{ \calh^i_{U_\epsilon}\otimes \calh^{i\prime}_{U'_\epsilon} \hookrightarrow \calh\ .}
However, again this structure is restricted by the locality bound, so that if the combined states in the two regions produce an $R(E)$ bigger than the separation between the regions, a simple description of the inclusion fails.  Of course, one of the questions for the future is the behavior of the network in these limiting, strong-gravity, regimes.\foot{A possibly related structure has been described in \refs{\BGH}.}

Such a network of Hilbert space inclusion maps clearly serves as a possible gravitational alternative to the LQFT net of subalgebras, providing a proposed candidate for the gravitational substrate.

Another open question is whether the network structure just described will extend to the full non-perturbative theory.  To see one of the issues, note that states or operators satisfying the constraints may be translated by any finite distance by operators that have only asymptotic support.  This is because, given a solution of the gravitational constraints, the momentum operators can be written in terms of integrals of the metric perturbation at infinity; a similar argument can also be given for anti de Sitter space\refs{\GiKi}.  So, for a gauge-invariant operator $\Phi(x)$ such as we have described, 
\eqn\phitrans{e^{-ia^\mu P_\mu} \Phi(x) e^{ia^\mu P_\mu} = \Phi(x+a)\ ,}
with $P_\mu$ an integral involving the metric at infinity.
This expression requires the full exponential, and all orders in the dressing, in order to be true\DoGithree; moreover, the momentum operators scale as $1/\kappa$ when written in terms of the metric perturbation $h_{\mu\nu}$.  And, of course if we were trying to measure the field through a gauge-invariant operator that is distant from $x$, that operator would be likewise translated through such conjugation.  So, plausibly the structure does extend, but that remains for further investigation.

\newsec{Contrast with other approaches}

One widely-discussed recent approach to quantum gravity is that of ``ER=EPR," or, more generally the idea that spacetime is built up from entanglement\refs{\vanR,\MaSu}.  From the present viewpoint, this approach can be criticized.  First, as was noted, entanglement is something that is defined {\it relative to a subsystem decomposition.}  And, it is precisely spacetime, or the analogous structure of a gravitational substrate on the Hilbert space, that defines a subsystem decomposition to begin with. To see the first point, consider the classic example of entanglement -- an EPR pair, in a state $|\Phi^\pm\rangle = (|0\rangle |1\rangle \pm |1\rangle|0\rangle)/\sqrt 2$, with entanglement entropy $\ln 2$.  Of course, here it is implicit that we are working on a four-dimensional Hilbert space, spanned by states $|0\rangle|0\rangle$, $|0\rangle |1\rangle$, $|1\rangle|0\rangle$, $|1\rangle|1\rangle$ arising from the obvious subsystem division.  But, an equally good basis for the space, ignoring this subsystem division, is the set of states $|\Phi^\pm\rangle$ and $|\Psi^\pm\rangle = (|0\rangle |0\rangle \pm |1\rangle|1\rangle)/\sqrt 2$.  And, one can relabel these states\refs{\ZLL} as $|\chi\rangle |\lambda\rangle$, with $\chi=\Phi,\Psi$ and $\lambda=+,-$, to define a different subsystem decomposition.  With respect to the latter decomposition, the EPR state is unentangled, and has entropy zero.

So, subsystems -- which in LQFT arise from spacetime and in quantum gravity must arise from analogous mathematical structure on $\calh$ -- are {\it prior} to the definition of entanglement.  From this perspective, it is hard to see how spacetime could emerge from entanglement.  

In a more familiar approach,  spacetime structure -- and presumably its quantum analog -- {\it implies} entanglement.  For example, in LQFT and in typical low-energy states, if we consider neighboring regions, the states are highly entangled between the regions.  Here, spacetime proximity implies entanglement.  However, the converse is not true.  We can have an entangled EPR pair of widely-separated particles.  We can observe that they are widely separated relative to the structure of the underlying manifold.  Or, this is also reflected in the structure of the Hamiltonian -- the Hamiltonian will not permit signaling (information transfer) between the particles without a long time delay.  Put differently, merely entangling distant electrons doesn't mean that we introduce couplings in the Hamiltonian that allow them to directly interact; in fact the localization provided by the subsystem decomposition is hardwired into the Hamiltonian, if it is local.

In short, entanglement is necessary, but not sufficient, for spacetime proximity.  This is explained through interplay of the subsystem structure and the Hamiltonian.  In LQFT localization is built into the theory in the algebraic structure, and is then reflected in the structure of the Hamiltonian; it is not emergent.  One expects similar statements to arise in the full quantum theory, once the underlying subsystem structure of the gravitational substrate is provided.

More specifically, Carroll and collaborators \refs{\CCM\CaCa-\CaSi} have also recently been pursuing the idea of describing quantum gravity in terms of structure on Hilbert space.  However, there are some notable differences from the present paper in how they approach the problem of finding the mathematical structure of the theory.  First, they assume subsystem decompositions arising from tensor factorization of the bigger Hilbert space, which this paper has argued is seemingly 
problematic in a theory that matches onto LQFT plus gravity in the correspondence limit. (Similar comments apply to \refs{\HSTone,\HSTrev}.) Then there is the question of {\it how} such a subsystem decomposition is specified.  They suggest that one resorts to the work of \refs{\CPR}, which shows that a given Hamiltonian is not typically local with respect to a tensor product structure, but if it is, the tensor product structure is typically unique.  Thus, one approach to specifying this product structure is simply to specify the Hamiltonian.\foot{Of course there is an inherent problem in this approach for closed universes, as the Hamiltonian vanises.}  

It is not clear that these statements extend to the case of infinite-dimensional Hilbert spaces, or to a case where the localization structure is not of the precise form arising in tensor product factorizations of the Hilbert space; \refs{\CPR} points out some of the questions if one only has an algebraic net of observables, and we have found gravity to be more subtle still.  But, beyond that, it seems more plausible and economical if a theory is specified by {\it first} specifying the underlying mathematical structure providing localization -- whether a tensor product structure or a more general structure such as a gravitational substrate as described above -- and {\it then} describing a Hamiltonian that respects that structure.  This is analogous to how other quantum theories -- from lattice models to field theories based on an underlying manifold -- are typically constructed in physics.  

In addition, refs.~\refs{\CCM\CaCa-\CaSi} advocate that spacetime geometry arises from entanglement.  However, once one has specified the Hamiltonian, in their approach, one has already specified a notion of locality.  Put  succinctly, local Hamiltonians can produce nonlocal entanglement through evolution, so there is an inherent conflict in then using entanglement to characterize localization.  

A related question is that of disentangling the degrees of freedom on two sides of a codimension-one boundary between spatial regions, denoted, say, ``left" and ``right"  -- such as the two sides of Rindler space, or the two sides of an Einstein-Rosen bridge which has been associated to a thermofield double\refs{\Mald}.  Naively, one might expect that the Hilbert space factorizes as $\calh= \calh_L\otimes \calh_R$, and then by considering sufficiently entangled states, we ``build up" the connection between the left and right regions.  However, at least if one considers states governed by a hamiltonian of LQFT, $\calh$  and $ \calh_L\otimes \calh_R$ should not be thought of as the same Hilbert space; states of the latter have infinite energy.  This connects directly to the type-III property described above -- infinite entanglement must be broken to write a state as a simple left/right product.  Perhaps, if gravity led to the modification of introducing extra finite-energy UV degrees of freedom\refs{\Harl,\DoFr}, such a separation could be found.\foot{This is also connected to puzzles about Wilson line operators running between such regions\refs{\GuJa}, which also seem to indicate that factorization is problematic, or that such operators are state-dependent\refs{\PaRa}.}  However, at present there doesn't appear to be a compelling reason for such degrees of freedom to exist, and rather quantum behavior of gravity seems to indicate modifications to subsystem structure along the lines described.

Work related to ER=EPR\refs{\BRSSZ,\Sussfall} has also argued for an important role of complexity in characterizing properties of gravity. Complexity of a state can be defined in terms of the number of quantum operations needed to transition to that state.  However, it is assumed that only certain quantum operations are allowed, and the definition of the allowed operations relies on the assumption of a particular subsystem decomposition.  If arbitrary unitaries are allowed, it is trivial to map any state onto any other state in a single step.

The preceding discussion also connects to the proposal that quantum information, say about the internal state of a black hole, is contained in ``soft hair\refs{\HPS}."  While the story of soft hair is a pretty one, the existence of split structure \splstr\ at leading order in $\kappa$ appears to argue against a role for this proposal\DoGithree, which  involves the asymptotic weak-field limit thus the leading linear-order perturbation $h_{\mu\nu}$.  The reason is that we have found that a given matter distribution may be dressed with a gravitational field that only depends on its Poincar\'e charges.  Of course other dressings exist, corresponding to different configurations of soft hair, but those simply correspond to superposing different graviton radiation fields on the dressed matter configuration.  The information in the soft hair of \refs{\HPS} -- except that of the Poincar\'e charges -- is then independent of the information in the matter.  However, the question raised above regarding whether higher-order correlators of operators can measure more details of the matter state doesn't yet rule out the possibility that   there could be some ``higher-order gravitational hair" that is sensitive to other details of the matter configuration.  This question requires more detailed analysis.

\newsec{Outlook}

This paper has assumed that a final quantum theory of gravity should be a theory respecting sufficiently general postulates of quantum mechanics, and that the problem of defining the theory is then to find the appropriate mathematical structure on Hilbert space.  A subsystem structure appears essential, to address concerns that were expressed well by Einstein\Eins: ``if one renounces the assumption that what is present in different parts of space has an independent, real existence, then I do not at all see what physics is supposed to describe. For what is thought to be a `system' is, after all, just conventional, and I do not see how one is supposed to divide up the world objectively so that one can make statements about the parts."  While in local quantum field theory such structure is induced from the underlying spacetime manifold, gravity appears to behave differently.  If, as described, weak gravity is taken as a guide in the weak-field, correspondence limit, 
 the gauge symmetry of gravity indicates a different notion of localization than in field theory.  This paper has outlined some first modest steps towards finding a mathematical structure, or gravitational substrate, that implements that localization structure, using these important clues from weak gravity.  This has been based on a perturbative notion of separability into subsystems that arises from gravitational split structure\DoGithree.  We thus see important aspects of the perturbative structure of the theory, but a big question is how these fit into a more complete nonperturbative structure

One may also comment that if large-$N$ super-Yang Mills does indeed provide a quantum theory of gravity through the AdS/CFT correspondence, this can plausibly be cast in such a framework.  Specifically, the former theory does have a Hilbert-space description, and so part of the problem of understanding the hypothesized correspondence is to understand the mathematical structure on this that gives a description of bulk subsystems.  At the perturbative level in gravity, subsystems in AdS have a similar structure to that in this paper, as has been preliminarily investigated in \refs{\GiKi}.

Next steps include studying generalization of this work beyond leading perturbative order, and ultimately into a structure appropriate to the fully nonperturbative theory.  

A closely related problem is that of determining the form of the evolution law based on such a structure.  Indeed, in a general quantum theory we can think of there being different aspects to locality.  The first is the notion we have been exploring of {\it localization} -- such as provided by tensor factorization of the Hilbert space or other subsystem structure like has been described.  The second is the notion of {\it local propagation} (or principle of local action\Eins), which is a statement restricting  the speed at which the Hamiltonian can transmit information between distant subsystems.  Of course, in LQFT these are related by the Poincar\'e symmetries of the theory, such that localization of spacelike-separated subsystems matches the statement that information cannot be transferred between these subsystems; plausibly there is a similar relation in the more complete gravitational theory.  

In order to respect quantum mechanics, the evolution law must be unitary -- {\it e.g.} yielding a unitary S-matrix in the context of states corresponding to asymptotically flat boundary conditions.  Given the challenges of describing unitary quantum black hole (BH) evolution, this appears to be an important constraint.  As a reminder, in an {\it approximate} description of black holes, the Hilbert space factorizes as
\eqn\bhprod{\calh_{BH}\otimes \calh_{env}\ ,}
corresponding to states of the BH and environment.  Then, if the Hamiltonian does not allow transfer of information from $\calh_{BH}$ to  $\calh_{env}$, and if 
$\calh_{BH}$ disappears as the black hole evaporates, as LQFT appears to indicate,  unitarity is violated.

This paper has reviewed arguments that a factorization \bhprod\ is not quite correct when one accounts for properties of gravity.  However, subsystem structure that replaces \bhprod\ has also been described.  An important question is whether the gravitational modifications to \bhprod\ that we have found are sufficient to resolve the unitarity problem -- if nonlocality of quantized GR resolves the unitarity crises, we may see it perturbatively here.  For example, while it has been noted that the leading-order description of standard dressing appears to rule out the suggested\refs{\HPS} role of soft quantum hair, it may be that a different higher-order effect leads to information about the internal state of a BH being accessible from outside.

While this remains to be explored more completely, we have not seen an indication that the modifications to \bhprod\ are sufficient to restore unitarity.  If perturbative modifications to subsystem structure, {\it i.e.} localization, in gravity do not directly resolve the unitarity problem, it appears that this provides an additional important clue about the nature of evolution -- plausibly the existence of additional couplings between subsystems that go beyond quantized GR.  The approach to modeling unitary evolution of BHs that has been proposed in \refs{\SGmodels,\BHQIUE,\NVNL\NVNLT-\NVNLpost} has been to parameterize such couplings, of a form that doesn't violate our correspondence constraints.  If this is the case, the necessity of consistent quantum evolution of BHs is providing us significant additional information about the nonperturbative structure of the theory.

\bigskip\bigskip\centerline{{\bf Acknowledgments}}\nobreak

This material is based upon work supported in part by the U.S. Department of Energy, Office of Science, under Award Number {DE-SC}0011702.  I thank J. Hartle for useful conversations and comments on a draft of this paper.

\listrefs
\end